\documentstyle[twocolumn,epsfig,prl,aps]{revtex}
\begin{document} 
\twocolumn[\hsize\textwidth\columnwidth\hsize\csname@twocolumnfalse\endcsname

\title{Atomic topology and radial distribution functions of {\it a}-SiN$_x$ 
alloys. {\it Ab initio} simulations
}
\author{Fernando Alvarez and Ariel~A.~Valladares$^{*}$
}
\address{Instituto de Investigaciones en Materiales, UNAM, Apartado Postal 
70-360, M\'exico D.F., 04510, MEXICO\\
}
\date{06 June 2001}

\maketitle

\begin{abstract}

We report a new approach to simulate amorphous networks of covalently bonded 
materials that leads to excellent radial distribution functions and realistic 
atomic arrangements.  We apply it to generate the first {\it ab initio} 
structures of nitrogen-doped silicon, {\it a}-SiN$_x$, for thirteen values 
of $x$ from 0 to the nearly stoichiometric composition of $x$=1.29, using the 
Harris functional and thermally amorphizised, periodically continued, 
diamond-like cells with 64 atoms.  Partial radial features are reported for 
the first time and the total radial distribution functions agree very well 
with the few existing experiments.  Our results should stimulate further 
experimental and theoretical studies in amorphous covalent materials.

\smallskip


\smallskip

\end{abstract}

]

\section{Introduction}

The properties of amorphous silicon-nitrogen alloys have attracted a great deal of 
attention in the last decade; {\it a}-SiN$_x$ has electrical, optical and mechanical 
features useful in a variety of industrial applications and the strong covalency of 
their bonding makes them the prototype of covalent materials. Their optical gaps depend 
strongly on the nitrogen content $x$ for $0\leq x\leq1.33$ so they can be tuned to 
fit specific needs in solar cells; their total and partial radial distribution 
functions (RDFs) are practically unknown, except for the stoichiometric 
content.  Some semiempirical studies have been done on their electronic structure, 
optical gaps and RDFs and a first-principles approach has been used on an {\it ad 
hoc}-generated amorphous structure.  Therefore, any {\it ab initio} approach that 
adequately generates, describes and predicts features of {\it a}-SiN$_x$ may have a 
wider applicability to deal with other covalently-bonded amorphous solids; in particular, 
systems like {\it a}-GeN$_x$ should be amenable to our approach \cite{fazzio}.

Recently \cite{valla} we carried out {\it ab initio} Harris-functional-based studies 
of the atomic and electronic structure of pure and hydrogenated amorphous silicon, 
using 64 silicon atom cells plus hydrogens that diffuse within the cells. We now apply 
these methods to amorphous silicon-nitrogen alloys to test 
the adequacy and the predictive powers of our approach for covalent materials. To 
the best of our knowledge, these are the first {\it ab initio} thermally generated 
amorphous networks where studies of the atomic topology of {\it a}-SiN$_x$ are 
carried out; the content $x$ is in the range $0\leq x\leq 1.29$ where 
$x = y/(64-y)$ and $y$ is the number of nitrogen atoms.

The experimental and theoretical activity prior to 1990 is well documented in a 
paper by Robertson where pertinent references can be found \cite{robertson}.  
In this work Robertson reports his semiempirical tight-binding calculations for 
the optical gaps and also reports several experimental gaps for hydrogenated and 
non-hydrogenated {\it a}-SiN$_x$ alloys.  Bethe lattice calculations have been 
done by Mart\'{\i}n-Moreno {\it et al.} \cite{martin} and by San-Fabi\'an 
{\it et al.} \cite{san} also using semiempirical parameters, whereas Ordej\'on 
and Yndur\'ain \cite{orde} in a very nice work do non-parameterized calculations 
of {\it a}-SiN$_x$ where the equilibrium positions of Si and N atoms in clusters 
are ported to the alloy network constructed {\it ad hoc}. They obtain a wealth of 
information including optical gaps, however tetrahedral coordination of the 
silicon atoms and threefold planar coordination of the nitrogen atoms is 
{\it assumed} with interatomic distances of 2.33 \AA\ for Si-Si and 1.74-1.76 \AA\ 
for Si-N.  A general characteristic of these calculations/simulations is that gap 
states, when considered, are introduced either by hand, progressively replacing Si 
by N, or by the algorithms that generate the random networks, {\it unlike} the 
procedure reported here. Recent semiempirical classical simulations by de Brito 
{\it et al.} \cite{brito} produced total RDFs and average nearest-neighbor results 
that are the subject of comparison with ours.

On the experimental side, as early as 1976 Voskoboynikov {\it et al.} \cite{vos} 
studied some RDFs and optical gaps of silicon-rich silicon-nitrogen films 
as a function of the gas ratio.  It was then observed that the gaps increase as a 
function of the nitrogen content; the films seemed to contain hydrogen and 
large clusters of silicon.  However, reliable experimental RDFs are scarce 
\cite{hase} and, apparently, only total ones for the stoichiometric 
amorphous composition exist \cite{aiya}, where a decomposition of the second peak of 
the total RDF into its partial contributions is also carried out.

It is clear that the atomic topology also determines the electronic properties of 
the amorphous samples, and therefore any understanding of the RDFs and the atomic 
distribution in the random networks is relevant in the characterization of the 
electronic and optical properties of these materials. In what follows we report 
the generation of random networks for amorphous silicon nitrogen alloys that leads 
to RDFs in good agreement with what is known experimentally and predict atomic 
structures for a variety of nitrogen contents. Electronic gap states are found 
in these structures that shall be dealt with in a future publication.

\section{Method}

Our 26 amorphous samples of {\it a}-SiN$_x$ were generated with {\it FastStructure} 
\cite{fast}, a DFT code based on the Harris functional produced by {\it Molecular 
Simulations, Inc.}.  The optimization techniques use a fast force generator to allow 
simulated annealing/molecular dynamics studies with quantum force calculations 
\cite{harris2}.  The LDA parameterization invoked is that due to Vosko, Wilk and 
Nusair \cite{vosko}.  An all electron calculation is carried out, and a minimal 
basis set of atomic orbitals was chosen with a cutoff radius of 5 \AA\ for the 
amorphization and 3 \AA\ for the optimization.  The physical masses of nitrogen and 
silicon are used throughout and this allows us to see realistic randomizing 
processes of the atoms during the amorphization of the supercell.  Finally, the 
forces are calculated using rigorous formal derivatives of the expression for the 
energy in the Harris functional \cite{lin}.

In order to test the adequacy of calculations carried out with {\it FastStructure}  
we used it to obtain the size of the crystalline cell of 
$\beta$-Si$_3$N$_4$ that minimizes the energy at the $\Gamma$-point.  
Fig. 1 shows the results of such calculation.  The experimental crystalline volumen 
is given by 145.920 \AA$^3$ \cite{wild} whereas the calculated volumen is 
146.797 \AA$^3$; a deviation of 0.6\%.  For this reason we feel cautiously 
optimistic about the use of {\it FastStructure} to generate random networks of 
silicon-nitrogen alloys.

Since it has become increasingly clear that quenching from a melt generates 
undesirable structures \cite{brito} we took a different path \cite{valla}.   We 
amorphisized the crystalline diamond structures with a total of 64 atoms ($(64-y)$ 
silicons and $y$ substitutional nitrogens) in the cell by slowly heating it, 
linearly, from room temperature to just below the corresponding melting point for 
each $x$, and then slowly cooling it to 0 K.  To determine the melting temperatures 
for each $x$ we linearly interpolated between the pure silicon value and the 
stoichiometric compound, $x=4/3=1.33$, and then remained below these temperatures 
(Table I).  Since the time step was the same for all runs, 6 fs, and the melting 
temperatures increased with $x$, the heating/cooling rate varied from 
$2.30\times10^{15}$K/s for pure silicon, to $3.11\times10^{15}$K/s for $x=1.29$.  
The atoms were allowed to move freely within each cell whose volume was determined 
by the corresponding density and content, Table I.  The densities were taken from 
the experimental results of Guraya {\it et al.} \cite{hase}.  Once this first stage 
was completed, we subjected each cell to annealing cycles with intermediate quenching 
processes.  Finally the samples were energy-optimized to make sure the final 
structures would be those of a local energy minimum.

\begin{center}
\begin{tabular}{|c|c|c|c|}
\multicolumn{4}{c}{\bf TABLE I}\\ 
\multicolumn{4}{c}{Contents, Melting Temperatures and Densities } \\
\multicolumn{4}{c}{for {\it a}-SiN$_x$}\\      \hline \hline
\ \ Sample \ \      &$x$         &Melting     &Density     \\
                    &            &Temp. (K)   &(g/cc)      \\ [0.05in]
\hline \hline 
Si$_{64}$N$_{0}$    &\ \ 0.000 \ \    &1680    &\ 2.329 \  \\  
Si$_{59}$N$_{5}$    &0.085            &1747    &2.435      \\ 
Si$_{54}$N$_{10}$   &0.185            &1814    &2.512      \\
Si$_{49}$N$_{15}$   &0.306            &1881    &2.600      \\
Si$_{44}$N$_{20}$   &0.455            &1948    &2.694      \\
Si$_{39}$N$_{25}$   &0.641            &2015    &2.803      \\
Si$_{34}$N$_{30}$   &0.882            &2082    &2.931      \\ 
Si$_{33}$N$_{31}$   &0.939            &2095    &2.957      \\ 
Si$_{32}$N$_{32}$   &1.000            &2108    &2.988      \\ 
Si$_{31}$N$_{33}$   &1.065            &2122    &3.017      \\ 
Si$_{30}$N$_{34}$   &1.133            &2136    &3.048      \\ 
Si$_{29}$N$_{35}$   &1.207            &2149    &3.081      \\ 
Si$_{28}$N$_{36}$   &1.286            &2162    &3.115      \\ [0.05in] 
\hline \hline
\end{tabular}
\end{center}

It should be kept in mind that our objective is always to generate realistic 
amorphous structures of {\it a}-SiN$_x$ and not, in any way, to mimic the 
experimental processes used to produce these alloys.

\section{Results and discussion}

We performed two runs for each $x$ value and from $x=0.882$ on 
({\it a}-Si$_{34}$N$_{30}$) the number of nitrogens was increased one at a time 
to be able to map the interesting processes that occur for these contents 
(percolation of the Si-Si bonds, widening of the optical gaps, etc).  Once the atomic 
structures were obtained, we calculated the corresponding total and partial RDFs for 
each of the 26 runs and averaged them by corresponding pairs.  Of those 13 
averaged plots we are reporting here total and partial RDFs for pure silicon, for the 
almost stoichiometric sample, {\it a}-Si$_{28}$N$_{36}$, $x=1.29$, and two intermediate 
ones: {\it a}-Si$_{44}$N$_{20}$, $x=0.46$ and {\it a}-Si$_{32}$N$_{32}$, $x=1.00$. 
Fig 2a for the pure amorphous silicon sample shows our results and the upper and lower 
bounds of the available experimental data \cite{valla}; the agreement is good since 
our RDF falls within these bounds and the four experimental peaks are correctly 
reproduced by our simulations. Figs. 2b to 2d show the variation of the partial RDFs 
for Si-Si, Si-N and N-N as a function of content and their contribution to the total 
RDF. In Fig. 2d the composition of the second peak of the total RDF for the nearly 
{\it a}-Si$_3$N$_4$ can be observed and it agrees completely with experiment 
\cite{misa}, since it is formed by the average second-neighbor $\left<2n\right>$ 
contributions of mainly the N-N and Si-Si partials and to a lesser extent by the Si-N 
partial. The third peak is essentially due to the Si-N partial with a small contribution 
from the N-N partial but, as far as we know, there are no experimental results for
comparison.

Figs. 2 show that as the nitrogen content increases the first peak of the total RDF 
(1.85 \AA), which is due to the Si-N average nearest-neighbor ($\left<nn\right>$) 
contributions, increases systematically and the $\left<nn\right>$ Si-Si peak (2.45 \AA) 
decreases systematically. The third peak moves toward low $r$ as $x$ increases since 
the N-N contribution becomes more predominant at high content (3.25 \AA\ to 2.95 \AA). 
In our structures there are no $\left<nn\right>$ nitrogens since the content is below 
stoichiometry and nitrogens have a marked tendency to bind to silicons. For the nearly 
stoichiometric sample, $x=1.29$, the Si-Si $\left<nn\right>$ contribution to the total 
RDF has practically dissapeared and this implies that there is a nitrogen atom between 
every pair of silicons. This is borne out by the results presented in Figs. 3 where a 
study of the the average coordination numbers $\left<cn\right>$ in the 13 random 
networks is depicted. The following cutoff radii were used: Si-Si, 2.55 \AA; N-N, 3.35 
\AA; and Si-N, 2.15 \AA, which are the positions of the minima after the first peaks of 
the corresponding partials.  Fig. 3a shows the results of our simulations where it can be 
seen that the N-N plot flattens for $x\approx1.1$, the percolation threshold of Si-Si 
bonds \cite{martinez}; the Si-Si $\left<nn\right>$ go from 4 to practically 0. The Si-N 
graph refers to the $\left<nn\right>$ nitrogens around the silicon atoms and varies from 
0 to 4, whereas the N-Si refers to the $\left<nn\right>$ silicons around nitrogens and 
indicates that nitrogens immediately surround themselves with practically 3 Si, 
saturating its valence. The crossing of the Si-Si and Si-N plots at $x\approx0.7$, is in 
agreement with experiment (Davis {\it et al.} \cite{hase}). There is a crossing of the 
Si-Si, N-Si and N-N plots at $x\approx0.3$ and a crossing of N-Si and Si-N at 
$x\approx1.0$ which have been observed experimentally for {\it hydrogenated} alloys 
by Guraya {\it et al.} \cite{hase}, Fig. 3b. However, due to the presence of hydrogen 
a curvature appears in the $\left<cn\right>$ for Si-Si, Si-N and N-Si so, in order to 
compare our results to this experiment, we did the following. We carried out the sum of 
N-H plus N-Si from the experiment, the average total number of atoms that surround a N, 
N-$*$, and plotted it along with our N-Si; we also did the sum of the experimental Si-N, 
Si-Si and Si-H, the average total number of atoms that surround a Si, Si-$*$, and plotted 
that along with our sum of Si-N plus Si-Si. This is presented in Fig. 3c. It is clear 
that our predictions closely agree with the integrated experimental results and validate 
our approach. The discrepancies are most likely due to existing dangling and floating 
bonds.

Fig. 4 is the comparison of our {\it ab initio} results with the classical Monte 
Carlo simulations of de Brito {\it et al.} \cite{brito} where they used empirical 
potentials developed {\it a la} Tersoff for the interactions between Si and N. It is 
clear that although the positions of some peaks are reproduced in both simulations, the 
general behavior of the total RDFs only agree qualitatively. Comparison of each of these 
simulations with experiment is presented in Fig. 5 where agreements and discrepancies 
can be appreciated.

\section{Conclusions}

We have devised an {\it ab initio} approach for {\it a}-SiN$_x$ 
($0\leq x \leq 1.29$) that generates radial distribution functions in 
agreement with existing experimental results and realistic atomic structures 
where the simulated average coordination numbers and the experimental ones 
coincide. Total RDFs agree very well with experiment, where available, and 
partial RDFs show that the Si-Si $\left<nn\right>$ peak disappears as nitrogen increases 
indicating a tendency to form 6-atom arrangements as the content $x$ approaches 
the stoichiometric value. Experiment shows that for {\it a}-Si$_3$N$_4$ Si and N 
form closed rings, Si-N-Si-N-Si-N, typical of the crystalline $Si_3N_4$ structures. 
The growth of the Si-N peak as nitrogen increases bears out this behavior.  No 
nitrogen-nitrogen bonds, including molecular nitrogen, are observed in the final 
structures even though for $x>1$ the starting diamond structure {\it does} contain 
nitrogens next to one another.  For $x\approx 1.1$ the effects of the percolation 
threshold of the Si-Si bonds is observed in the N-N $\left<2n\right>$.  For $x\approx 0.7$ 
the Si-Si and Si-N neighbors are practically the same, as found experimentally. 
Also, Si-Si, N-Si an N-N are practically the same for $x\approx 0.3$ as are Si-N and 
N-Si for $x\approx 1.0$. The integrated experimental results and our simulation agree. 
The first prominent peak in the total RDF of the nearly stoichiometric 
sample is due to Si-N and an analysis of the second peak indicates that N-N, Si-Si 
and Si-N contribute to it, in agreement with experiment.  The third peak is mainly 
due to the Si-N, with a small contribution from the N-N; no experimental results 
exist for comparison.  Our approach, being {\it ab initio}, is of wider 
applicability than classical or semiempirical ones and should be relevant for the 
understanding of the physics of amorphous covalent materials.

\acknowledgments

AAV thanks DGAPA-UNAM for financing the project IN101798 and a sabbattical stay at 
MSI-San Diego, USA, where this work was begun.  FA thanks CONACyT for supporting his 
PhD studies.  This work was done on an Origin 2000 computer provided by DGSCA, UNAM.

\pagebreak

Figure1. Energy vs volume for crystalline beta-Si3N4 obtained
with FastStructure and experimental values taken from Wild  et al. 
(see text). The experimental and calculated volumes agree to within 0.6\%.

Figure2. Total and partial RDFs for (a) pure silicon; the light curves are the upper 
and lower experimental bounds (Ref. [2]); (b) a-Si44N20, x=0.46; 
(c) a-Si32N32, x=1.00; (d) the nearly stoichiometric 
sample a-Si28N36, x=1.29.

Figure3. Average coordination numbers <cn> as a function of x. 
(a) Our results. (b) Experimental results for hydrogenated alloys from Guraya 
et al.. (c) Comparison of the integrated results (see text). Lines are 
drawn as guides to the eye.

Figure4. Comparison of our results and those obtained by de Brito et al. 
(Ref. [7]) for the total RDF. The agreement is at best qualitative although the 
position of some of the peaks coincide.

Figure5. (a) Comparison of our simulations and experiment (Ref. [10]) for the 
stoichiometric sample. (b) Comparison of de Brito's simulations and experiment 
for the same sample.

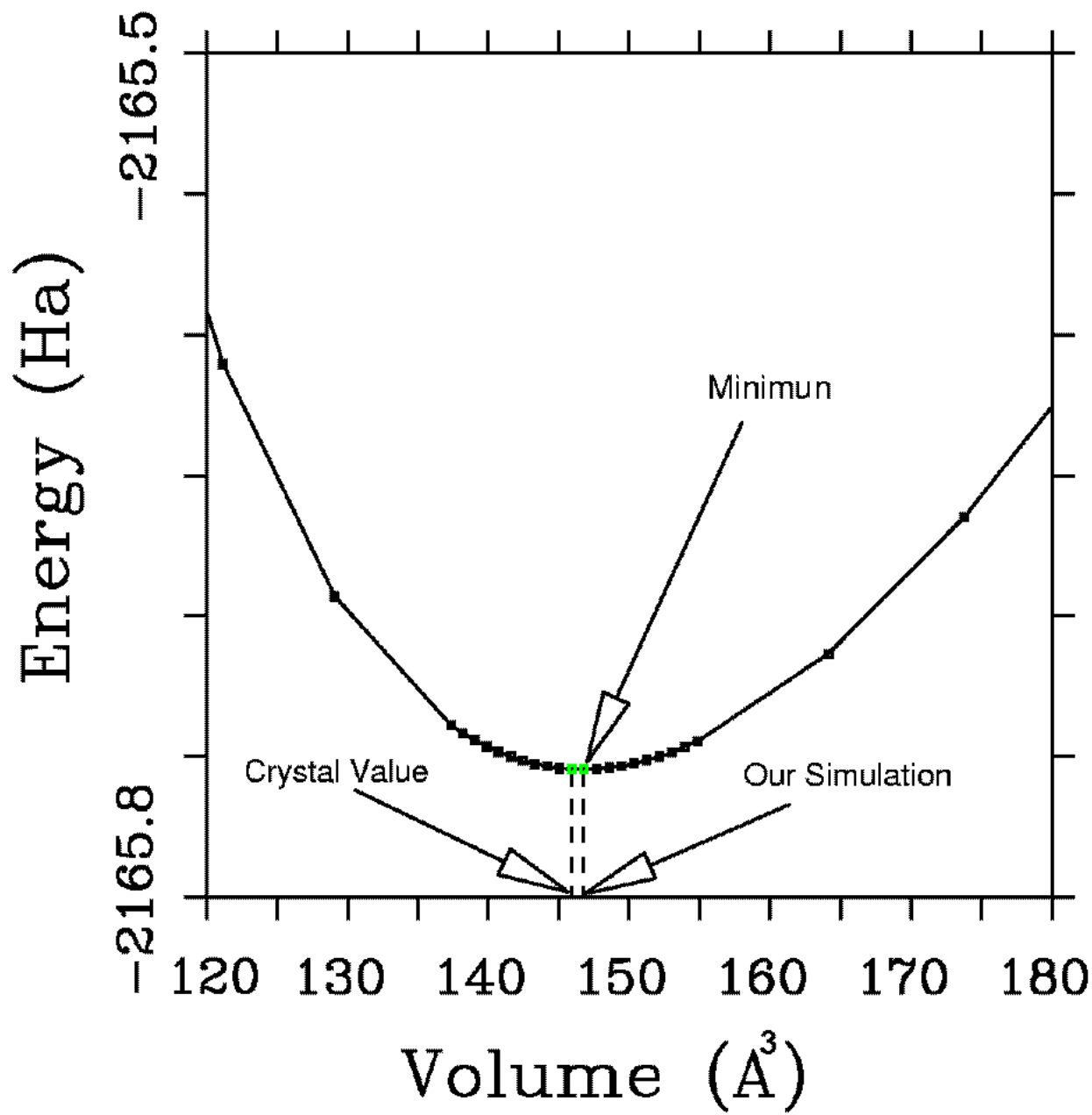
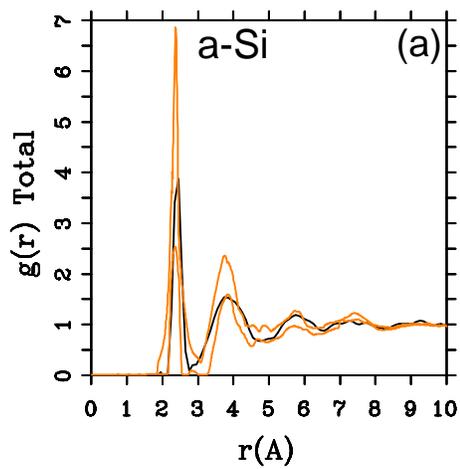
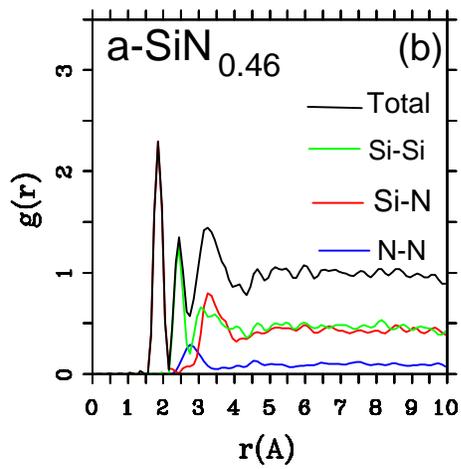
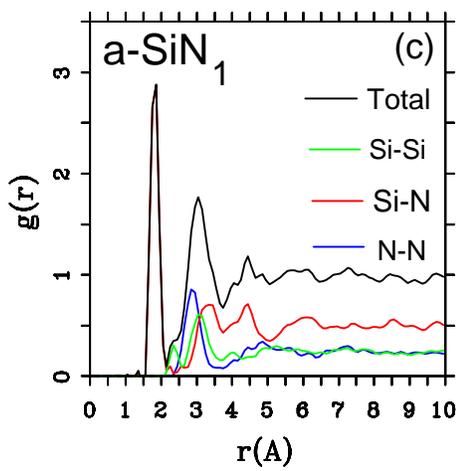
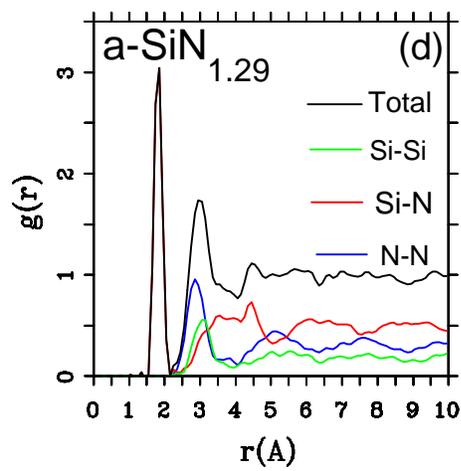
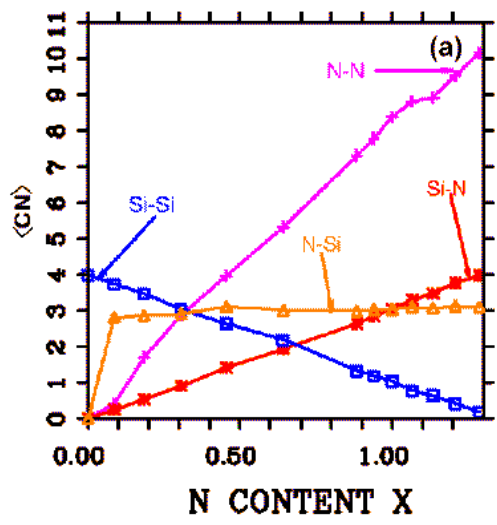
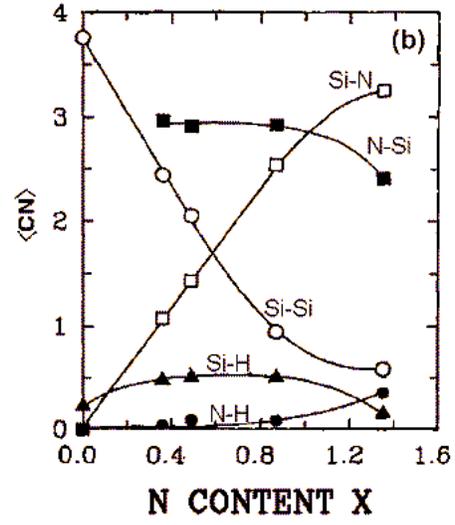
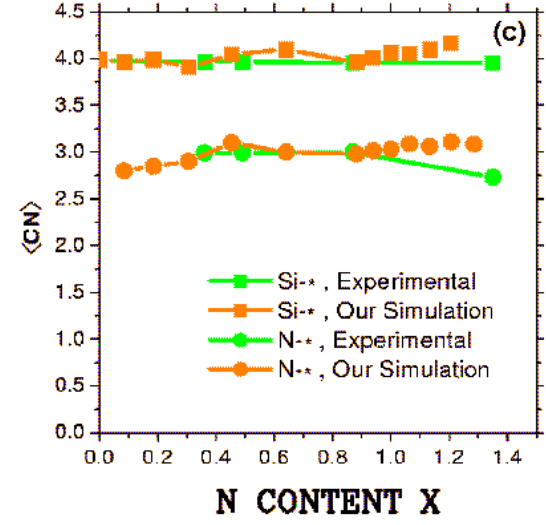
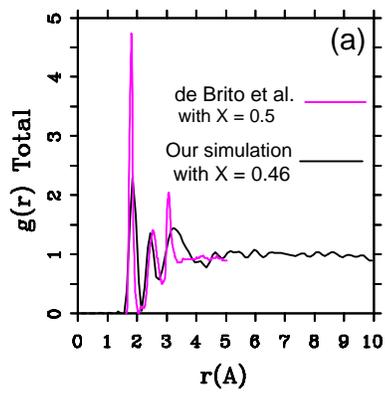
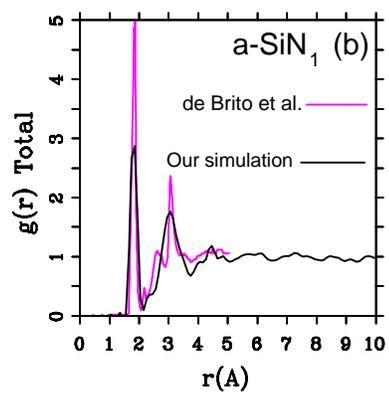
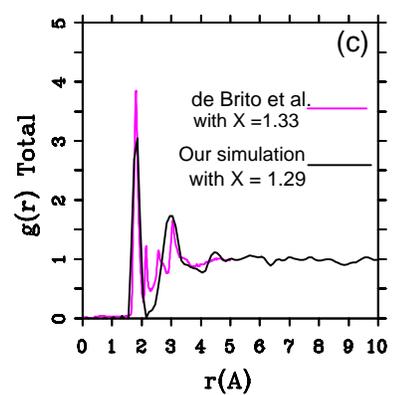
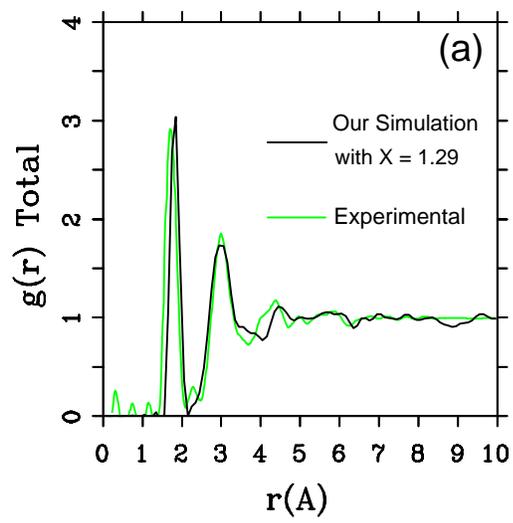
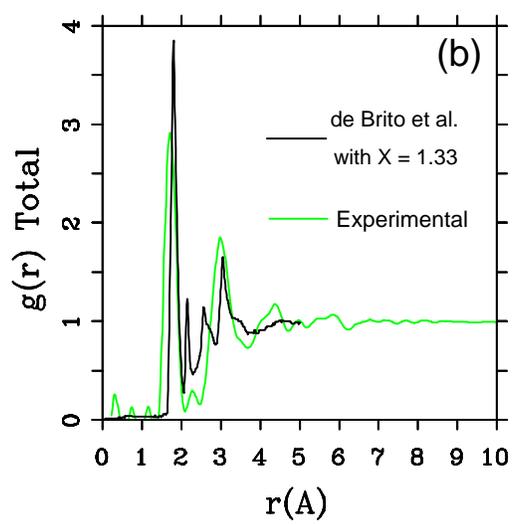
\end{document}